\documentclass[aps,prb,twocolumn]{revtex4}
\usepackage{graphicx}

\begin{document}
\title{Origin of Ferromagnetism and its pressure and doping dependence in Tl$_{2}$Mn$_{2}$O$_{7}$}
\author{T. Saha-Dasgupta$^{1}$, Molly De Raychaudhury$^{1}$ and D. D. Sarma$^{2,}$$^*$}
\affiliation{$^1$ S.N. Bose National Centre for Basic Sciences,
Kolkata 700098, India} \affiliation{$^2$ Solid State and
Structural Chemistry Unit, Indian Institute of Science, Bangalore
- 560012, India}
\pacs{75.47.Gk, 71.20.-b, 71.70.Gm}
\date{\today}

\begin{abstract}
Using NMTO-{\it downfolding} technique, we explore and establish
the origin of ferromagnetism in the pyrochlore system,
Tl$_{2}$Mn$_{2}$O$_{7}$. It is found to be driven by hybridization
induced spin-polarization of
the delocalized charge carriers derived from Tl-$s$ and O-$p$
states. The mean-field estimate of the ferromagnetic transition
temperature, T$_c$, estimated using computed exchange integrals
are found to be in good agreement with the measurements. We find
an enhancement of T$_{c}$ for moderate doping with nonmagnetic Sb
and a suppression of T$_{c}$ upon application of pressure, both in
agreement with experimental findings.
\end{abstract}
\maketitle
\noindent
A range of compounds, mostly perovskites based on manganese oxide, have been
found with astoundingly large negative magnetoresistance (MR). The search is 
on for materials with better MR properties, either in the form of colossal
magnetoresistance (CMR) or tunneling magnetoresistance (TMR).
Examples include systems such as double perovskites,\cite{double}
pyrochlores,\cite{pyro} FeCr$_{2}$S$_{4}$ and
Fe$_{0.5}$Cu$_{0.5}$Cr$_{2}$S$_{4}$ chalcospinels,\cite{chalco}
layered rare-earth iodide GdI$_{2}$.\cite{gdi} While work in 1950s
and 1960s uncovered\cite{dex} the role of double-exchange (DEX)
mechanism in providing a qualitative understanding of
stabilization of the ferromagnetic (FM) phase in perovskite
manganites, the underlying driving mechanism for the magnetic
order in diverse class of magnetoresistive materials need not be
the same. Our work\cite{prl-srfemo} on Sr$_{2}$FeMoO$_{6}$ in the 
past has shown that driving mechanism acting in Sr$_{2}$FeMoO$_{6}$ 
is neither the conventional double-exchange mechanism nor the 
super-exchange, instead a novel kinetic energy driven mechanism 
explains its unusual electronic and magnetic behavior. In the present
communication, we focus on a pyrochlore manganite
Tl$_{2}$Mn$_{2}$O$_{7}$, which has a number of contrasting
properties compared to perovskite manganites. For example,
Tl$_{2}$Mn$_{2}$O$_{7}$ does not have mixed Mn$^{3+}$-Mn$^{4+}$
valences, they do not exhibit Jahn-Teller distortions in the
MnO$_{6}$ octahedra, and in spite of similar CMR effect observed,
both the ferro- and para-magnetic phases show a metal-like
behavior.

Since the discovery of CMR effect in Tl$_{2}$Mn$_{2}$O$_{7}$, it
has been recognized as a class of compounds that does not fit
within the DEX framework. Nevertheless, no common consensus has
emerged concerning the driving mechanism in this interesting class
of compounds. Considering the fact that Mn-O-Mn bond angle in
Tl$_{2}$Mn$_{2}$O$_{7}$ is substantially reduced\cite{struc} from
180$^{o}$ to $\approx$ 133$^{o}$, a value that falls in the range
where a sign change of the exchange interaction from
antiferromagnetic to ferromagnetic is expected according to
Goodenough-Kanamori rule, \cite{gf} a ferromagnetic superexchange
picture was proposed originally.\cite{superex} Later on this
interpretation has been questioned in view of the fact that
ferromagnetic T$_c$ is enhanced by introduction of moderate amount
of nonmagnetic cation like Sb in the Mn sublattice\cite{substi}
and gets suppressed by application of pressure,\cite{pressure}
contrary to the expectation based on super-exchange. The extended
version of super-exchange idea\cite{super_new} has been used to
explain the ferromagnetism in low T$_c$ pyrochlore materials,
making the case of Tl$_{2}$Mn$_{2}$O$_{7}$ with a much larger
T$_c$ even more intriguing. Mishra and Satpathy proposed
\cite{satpathy}a mechanism which is a combination of negative
Hund's rule energy (J$_{H}$) driven double exchange mechanism,
antiferromagnetic super-exchange mechanism and an indirect
exchange mechanism, though lacking rigorous microscopic
justification for these mechanisms. In view of the enhancement of
the ferromagnetic coupling in Sb substituted
Tl$_{2}$Mn$_{2}$O$_{7}$, a more exotic scenario\cite{substi} has
also been suggested with antiferromagnetic (AFM) coupling between
nearest-neighbor(NN) Mn ions dominated by longer-ranged FM
interactions due to the frustration of the former in the
pyrochlore lattice.

Given the diversity of disparate mechanisms for ferromagnetism, we
considered it worthwhile to study the underlying electronic
structure model, responsible for magnetism within a rigorous and
microscopic {\it ab-initio} theory of Tl$_{2}$Mn$_{2}$O$_{7}$. For this purpose, we
have analyzed the electronic structure of Tl$_{2}$Mn$_{2}$O$_{7}$, pristine, doped
with Sb and under pressure, computed within the framework of local
spin density approximation (LSDA) of density functional theory
(DFT) in terms of muffin-tin orbital (MTO) based NMTO-{\it
downfolding} technique. We have also computed the magnetic
exchange interaction strengths from {\it ab-initio} DFT
calculations.


The pyrochlore structure of general stoichiometry
A$_{2}$B$_{2}$O$_{6}$O$^{'}$ can be described as two
interpenetrating networks. The smaller B cations (Mn) 
are octahedrally co-ordinated by O-type of oxygens,
with the BO$_{6}$ octahedra sharing corners to give rise to a
BO$_{3}$ network composition. The cage-like hole of this network
(cf. Fig.\ \ref{orbital}) contains the second network comprising
of A (Tl) and O$^{'}$-type oxygens forming
A-O$^{'}$ chains with a formula A$_{2}$O$^{'}$. Self-consistent
calculations were carried out within the framework of tight-binding 
LMTO\cite{tb-lmto}, for Tl$_{2}$Mn$_{2}$O$_{7}$ in
$Fd3m$ symmetry. The calculation yields a net spin moment of 2.90
$\mu_B$ per Mn atom, consistent with the the measured values (2.74
$\mu_B$ and 2.59 $\mu_B$\cite{pyro,magmom}) and the
previous band-structure calculations. \cite{satpathy,david} Each
Mn site with a radius of 1.2 $\AA$ contributes 2.57 $\mu_B$,
while each 1.0 $\AA$ O$^{'}$ sphere and each 1.5 $\AA$ Tl
sphere contribute 0.24 and  0.08 $\mu_B$,  respectively.

\begin{figure}
\includegraphics[width=8.5cm,keepaspectratio]{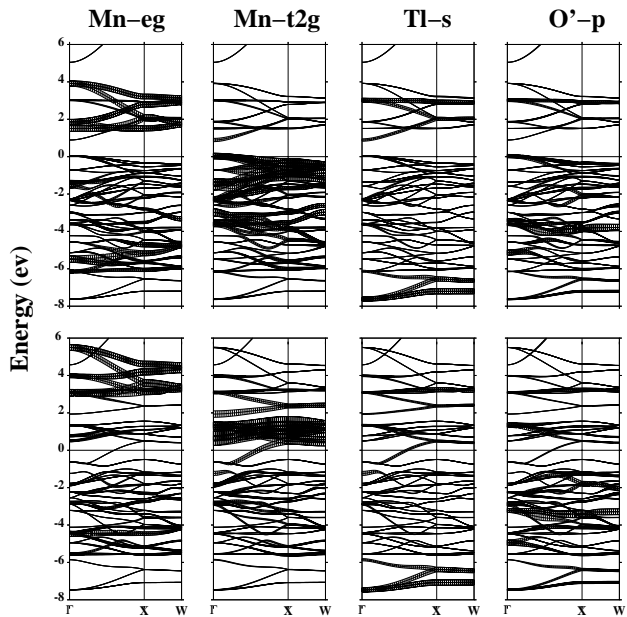}
\caption{Spin-polarized LDA band-structure of Tl$_{2}$Mn$_{2}$O$_{7}$. The upper panels correspond to up-spin channel
and the lower panels correspond to down-spin channel. The fatness associated with the bands in each
panel is proportional to the orbital character corresponding to that indicated on top of each panel.}
\label{fat}
\end{figure}
The well-studied \cite{satpathy, david} band-structure of
Tl$_{2}$Mn$_{2}$O$_{7}$, shown in  Fig.\ \ref{fat}, may be
summarized as follows. The $d$-$p$ hybridized band-structure
extends in an energy range from about - 6 eV to 6 eV with the zero
of energy set at the LSDA Fermi level, E$_F$. The occupied
band-structure in both spin channels are dominated by O-$p$
contributions (not shown in the figure). In the majority or up
spin channel the upper part of the occupied manifold, starting
from $\approx$ -2 eV is also strongly contributed by the
Mn-$t_{2g}$ states. The crystal field split Mn-$e_g$ states span
an energy range from 1.5 eV to 4 eV overlapping with the Tl-$s$
$-$ O$^{'}$-$p$ hybrid bands. The almost full Mn-$t_{2g}$ bands
separated by a gap from Mn-$e_g$ bands produce tiny hole pockets
in the majority spin channel formed out of three almost flat
bands. In the minority or down spin channel, the exchange split
Mn-$d$ bands are shifted up in energy by $\approx$ 2 eV, thereby
making the Mn-$d$ states close to empty in the down-spin channel.
The noticeable feature is the highly dispersive Mn-$t_{2g}$ band
hybridized with Tl-$s$ and O$^{'}$-$p$ that crosses E$_F$,
producing an electron pocket in the minority spin channel. The
unusually large Tl$-$O$^{'}$ hybridization produces Tl-O$^{'}$
bonding like states at the bottom of the spectrum at both spin
channels spanning the energy from $\approx$ -7 eV to  -6 eV.

The band-structure of Tl$_{2}$Mn$_{2}$O$_{7}$ shows a striking
similarity with that of Sr$_{2}$FeMoO$_{6}$ over the low to medium
energy scale. In case of Sr$_{2}$FeMoO$_{6}$, Fe-$d$ states are
completely full in the majority spin channel, while completely
empty Mo-$d$ states appear separated from E$_F$ by a gap of
$\approx$ 1 eV. In the minority spin channel the $t_{2g}$ manifold
of the crystal field split Fe-$d$ bands, strongly hybridized with
Mo-$t_{2g}$ (and O-$p$), cross the Fermi level. These
band-structure aspects suggest a strong renormalization of the
spin-splitting of the Mo-$t_{2g}$ bands (or more precisely the
Mo-t$_{2g}$ $-$ O-p hybridized states) over its bare value, since
Mo is usually {\it nonmagnetic} with 0.1 - 0.2 eV intrinsic
exchange splitting. This was explained\cite{prl-srfemo} in terms
of the large spin splitting at the Fe site and the presence of
substantial hopping between Fe and Mo sites. This situation may be
directly compared with that of Tl$_{2}$Mn$_{2}$O$_{7}$ with
half-filled $t_{2g}$ 3$d^{3}$ state of Mn in
Tl$_{2}$Mn$_{2}$O$_{7}$ playing the role of half-filled 3$d^{5}$
state of Fe in Sr$_{2}$FeMoO$_{6}$ and the hybridized Tl-O$^{'}$
state playing the role of delocalized Mo-O state in
Sr$_{2}$FeMoO$_{6}$.

\begin{figure}
\includegraphics[width=8.5cm,keepaspectratio]{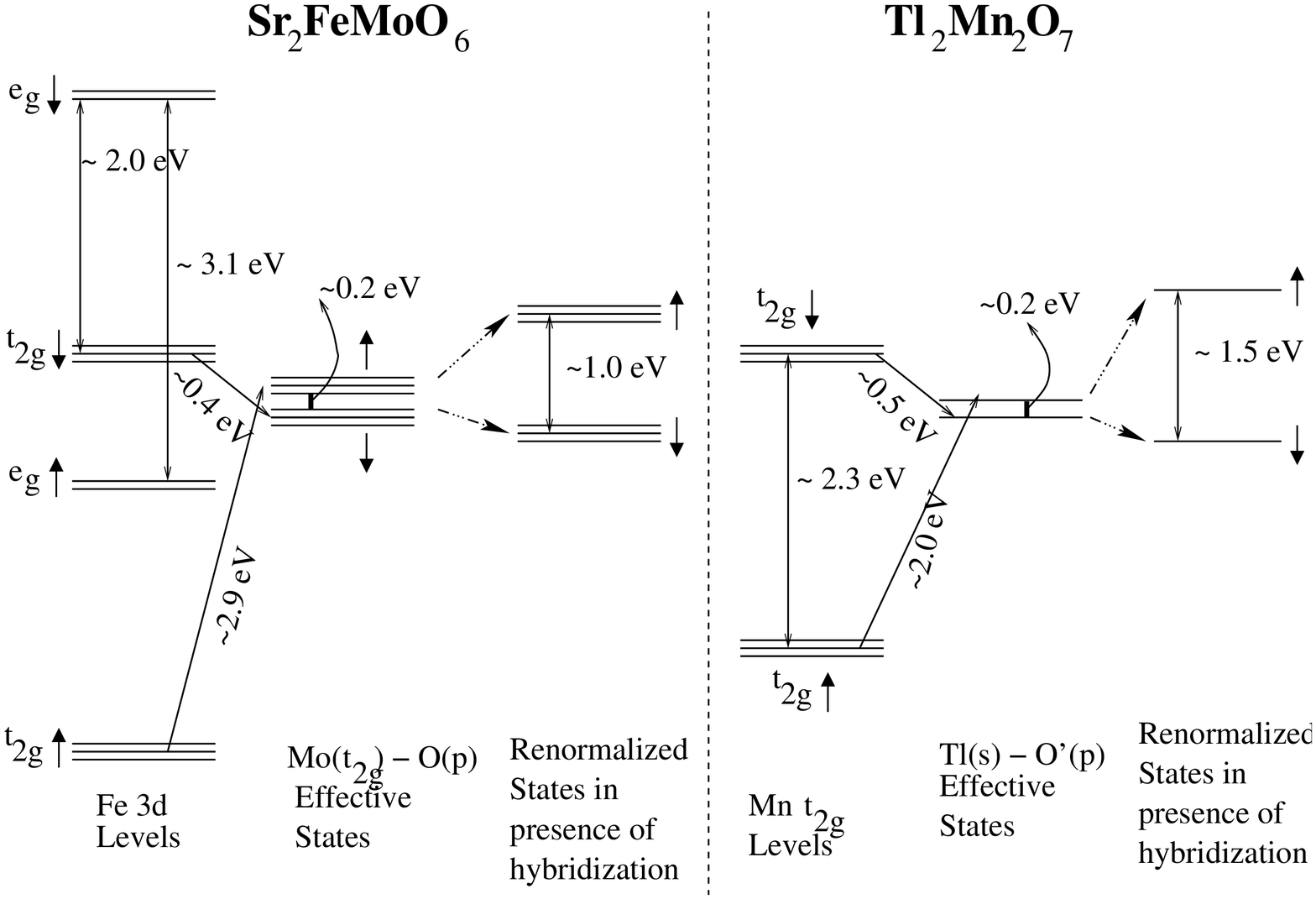}
\caption{Positioning of various energy levels as obtained by NMTO-{\it downfolding}
calculation before and after switching on the hybridization between the magnetic and nonmagnetic ions.}
\label{level}
\end{figure}

\begin{figure}
\includegraphics[width=8cm,keepaspectratio]{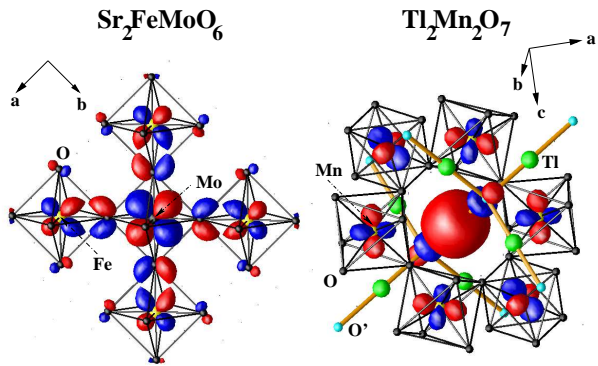}
\caption{(Color on-line) Effective Mo-t$_{2g}$ like and Tl-s like Wannier orbital corresponding to massively downfolded NMTO Hamiltonian
in the down spin channel. Shown are the orbital shapes (constant-amplitude surfaces) with lobes of opposite
signs colored as red and blue.}
\label{orbital}
\end{figure}

In the following we attempt to unravel the exchange mechanism by
application of NMTO-{\it downfolding} technique.\cite{nmto}
NMTO-{\it downfolding} technique provides a useful
way\cite{nmto} to derive few-orbital Hamiltonians starting
from a full LDA/LSDA Hamiltonian by integrating out degrees of
freedom that are not of interest. This procedure naturally takes
into account the renormalization effect due to the integrated-out
orbitals by defining energy-selective, effective orbitals which
serve as the Wannier or Wannier-like orbitals for the few-orbital
Hamiltonian. We employ the NMTO-{\it downfolding} technique to
construct the real-space Hamiltonian in the NMTO-Wannier function
basis for Sr$_{2}$FeMoO$_{6}$ and Tl$_{2}$Mn$_{2}$O$_{7}$. The O degrees of freedom for
Sr$_{2}$FeMoO$_{6}$ and O$^{'}$ degrees of freedom in case of Tl$_{2}$Mn$_{2}$O$_{7}$ are
downfolded to define effective Mo-O and Tl-O$^{'}$ states
respectively. The on-site matrix elements of these real-space
Hamiltonians give the estimate of various energy level positions
in absence of the hybridization between the localized magnetic (Fe
or Mn) and the delocalized non-magnetic (effective Mo-O or
effective Tl-O$^{'}$) states. As illustrated in Fig.\ \ref{level},
the essentially non-magnetic Mo-O or Tl-O$^{'}$ states with a tiny
exchange splitting ($\leq$ 0.2 eV)\cite{note_split} appear in
between the exchange split Fe-$d$ or Mn-$t_{2g}$ states
\cite{note}, respectively.
On switching on the hybridization, as depicted in the right hand
side of the level diagram in Fig.\ \ref{level}, the Mo-O or
Tl-O$^{'}$ levels get pushed up in the up-spin channel and pushed
down in the down-spin channel resulting into a large, renormalized
negative spin-polarization of the mobile electron due to purely
hopping interactions. The renormalized spin-splittings of the Mo-O
or Tl-O$^{'}$ states, after switching on the hybridization, have
been estimated by massive downfolding calculations. In these
calculations, only the Mo-$t_{2g}$ or Tl-s states have been kept
active and all other degrees of freedom, including Fe-$d$ or
Mn-$t_{2g}$, have been downfolded to take into account the
hybridization induced renormalization effect.  Fig.\ \ref{orbital}
shows the Wannier orbitals corresponding to massively downfolded
NMTO Hamiltonian in the down-spin channel for Sr$_{2}$FeMoO$_{6}$
and Tl$_{2}$Mn$_{2}$O$_{7}$. The central Mo site 
(Sr$_{2}$FeMoO$_{6}$) or Tl site (Tl$_{2}$Mn$_{2}$O$_{7}$) shows
the expected Mo-$t_{2g}$ or Tl-$s$ character, while the {\it tail}
of the orbital sitting at the Fe and O sites (Sr$_{2}$FeMoO$_{6}$) 
or Mn, O and O$^{'}$ sites (Tl$_{2}$Mn$_{2}$O$_{7}$) has appreciable 
weight shaped according to Fe-$t_{2g}$, O-$p$ or Mn-$t_{2g}$, O-$p$, 
O$^{'}$-$p$ symmetries
indicating the basic hybridization effect responsible for the
renormalization. This kinetic energy driven mechanism enforces a
particular spin orientation of the mobile carriers with respect of
that of the localized spin, thereby providing a mechanism of
ferromagnetic ordering at the localized spin sub-lattice. This is
a general mechanism and will take place whenever the nonmagnetic,
partially occupied level is placed within the exchange split
energy levels of the magnetic ion as was emphasized in ref.20.

The above exercise demonstrates beyond any doubt that the
hybridization induced negative spin-polarization mechanism similar
to Sr$_{2}$FeMoO$_{6}$ to be operative also in case of
Tl$_{2}$Mn$_{2}$O$_{7}$. The reason that in case of
Tl$_{2}$Mn$_{2}$O$_{7}$, Tl in contrast exhibits a net {\it
positive} moment as opposed to the ferrimagnetic spin alignment of
Fe and Mo states in Sr$_{2}$FeMoO$_{6}$, is the unusual covalency
of Tl-O$^{'}$ and mixing with Mn-O states which extends till the
bottom of the spectra.\cite{david}

We now turn to estimate quantitatively the exchange interaction
strengths, and therefore T$_c$.
We compute the exchange integrals by comparing the LSDA total
energies of different spin configurations to that of an effective
Heisenberg Hamiltonian constructed out of Mn$^{4+}$ spins. 
The network of Mn ions in Tl$_{2}$Mn$_{2}$O$_{7}$, as shown in Table I, is an
infinite three-dimensional lattice of corner-sharing tetrahedra.
Such a geometrical arrangement gives rise to a very high
degree of frustration for AFM, NN interactions.  
Due to this frustration, it is not possible to satisfy the antiferromagnetism
completely, and the chosen AFM configurations invariably have net
magnetic moments. The energy differences of all the AFM configurations relative to FM configuration
are positive, implying FM state as the stable ground state consistent
with experimental findings.
\begin{table}
\begin{tabular}{lc}
\begin{minipage}{3cm}
{\includegraphics[width=3cm,keepaspectratio]{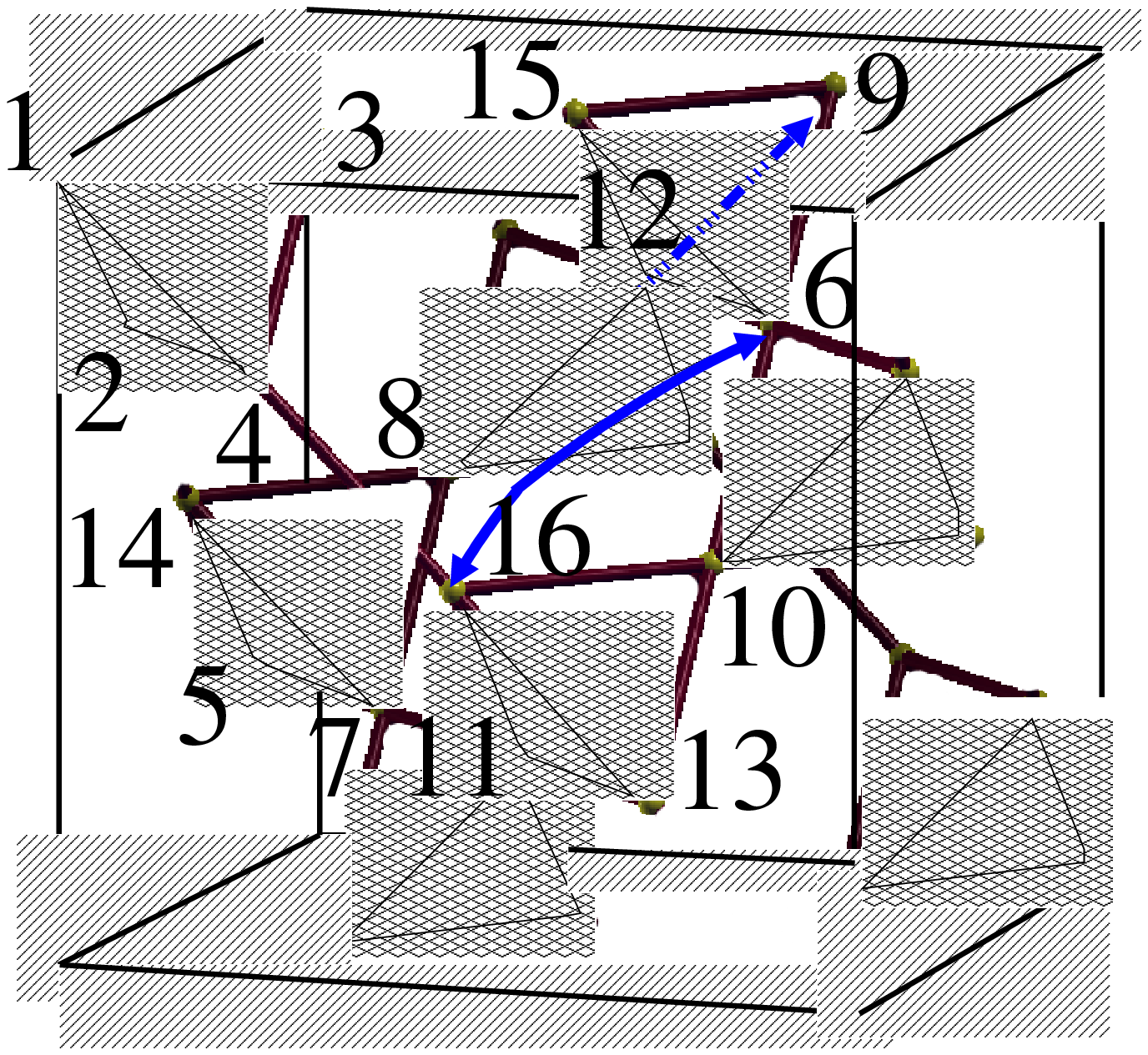}}
\end{minipage}
\hfill
&
\begin{minipage}{5.4cm}
{\frenchspacing {TABLE I: Magnetic configurations of the
Mn ions in the supercell for the states used to determine the magnetic interactions. The numbering of the Mn ions are as shown
in the picture. The last column gives the relative LSDA energies per Mn ion in meV. }}
\end{minipage}
\end{tabular}
\begin{tabular}{|cccccccccccccccccc|} \hline
& 1 & 2 & 3 & 4 & 5 & 6 & 7 & 8 & 9 & 10 & 11 & 12 & 13 & 14 & 15 & 16 & $\Delta$~E \\ \hline
FM  & + & + & + & + & + & + & + & + & + & + & + & + & + & + & + & + & 0 \\
AFM1 & + & - & + & + & + & - & + & + & + & - & + & + & + & - & + & + & 9.55\\
AFM2 & - & - & - & + & - & + & + & - & - & - & - & - & + & - & - & - & 13.98\\
AFM3 & + & + & - & - & + & + & - & - & + & + & - & - & + & + & - & - & 12.24 \\
AFM4 & - & - & + & + & - & + & + & + & + & + & - & - & + & - & - & - & 17.59 \\
AFM5 & - & - & + & + & - & - & + & + & - & + & - & + & + & - & + & - & 18.39 \\ \hline
\end{tabular}
\label{mag-inter}
\end{table}

The effective Heisenberg Hamiltonian, considering till 3NN interactions (six NN, 
twelve second NN and twelve third NN), can be written as,
\begin{equation}
H = J_{1} \sum_{nn} S_{i}.S_{j} + J_{2} \sum_{2nn} S_{i}.S_{j}
+ J_{3} \sum_{3nn} S_{i}.S_{j}
\end{equation}
where S$_{i}$ denotes the spin-3/2 operator corresponding
Mn$^{4+}$ states at site $i$, and $J_1$, $J_2$, $J_3$ denote the
NN, 2NN and 3NN magnetic exchange interaction strengths.
For the calculation of exchange couplings, the computed LSDA
energies were fitted to the mean-field Heisenberg model, which
contains Ising terms of the full Hamiltonian (1). A well-known
limitation of this approach is that in some cases the result
depends on the choice of spin configurations. To overcome this we
obtained seven different estimates of $J_1$, $J_2$ and $J_3$
employing various independent combinations of five different energies listed
in Table I. Our calculation gives
the average estimate of $J_1$, $J_2$, $J_3$ as -2.52 meV, -0.11
meV and 0.33 meV with a standard deviation of 0.24 meV, 0.06 meV
and 0.08 meV, respectively. Substituting the estimated values of
$J_1$, $J_2$, $J_3$ in the mean-field estimate of T$_{c}$:
$T^{mf}_{c} = \frac{S(S+1)J_{o}}{3 k_{B}}$, where $J_o$, the net
effective interaction is $6 J_1 + 12 J_2 + 12 J_3$, S=3/2 and
$k_B$ is the Boltzmann constant, gives a $T_c$ of 181 K. This is a
very reasonable estimate considering mean-field overestimation and
experimentally measured value of 142 K. \cite{pyro}

As mentioned in the beginning, magnetic properties of
Tl$_{2}$Mn$_{2}$O$_{7}$ exhibit interesting variations with
Sb-substitution and under pressure. We have addressed these issues
within the present approach. We mention here only the salient
points of our study, the details of the calculation will be
reported elsewhere.\cite{molly} To mimic the  Sb-substitution, we
have carried out calculations with eight formula unit supercells,
where 1 and 2 out of 16 Mn atoms are replaced by Sb, corresponding
to a doping level of 6.25 $\%$ and 12.5 $\%$, respectively. Due to
the slightly larger size of  Sb, lattice parameters and bond
lengths increase by less than 1$\%$ with no appreciable change in
bond angles.\cite{substi} The computed electronic structure shows
the overall band shapes to remain more or less same with
electron doping in the strongly hybridized
Tl-$s$--O-$p$--Mn-$t_{2g}$ conduction band in the down spin
channel. This causes a substantial rise in the density of states
of the conduction band at E$_F$, from 0.2 eV$^{-1}$ for the
undoped case to 1.0 eV$^{-1}$ and 1.4  eV$^{-1}$ for 6.25 $\%$ and
12.5 $\%$ doped cases, respectively. Within the framework of the
proposed mechanism, this is expected to lead to a strong
enhancement of the magnetic coupling.\cite{kanamori} This is
supported by our computed $J$'s from LSDA total energy
calculations, which necessarily includes the proposed kinetic
energy driven exchange. The mean field estimate of T$_c$ is found
to be 382 K and 420 K for  6.25 $\%$ and 12.5 $\%$
doping concentrations, respectively. Though the trend is in
agreement with the experimental observations,\cite{substi} the
enhancement is grossly overestimated, presumably due to clustering
and disordering effects, not taken into account in the
calculation.

Existing proposals for structural changes under pressure are
controversial. The first paper\cite{pressure} reported a
decrease of Mn-O-Mn bond angle, in addition to a decrease of bond
lengths. Subsequent measurements\cite{alonso_new} claim the
Mn-O-Mn bond angle to increase.
In absence of a clear consensus about the structural changes, we carried out two sets of
calculations. In the first set, the pressure applied is assumed to
be isotropic, so that only the bond lengths are assumed to
decrease. In the second set, we also changed the Mn-O-Mn bond
angle, taken according to Fig. 2 of ref.23. Using the first set of
calculations, the mean field estimate of T$_c$ derived from our
computed $J'$s, show a decrease of T$_c$ by 40 K upon 2 $\%$
change in bond lengths. In second set of calculations, where we
also changed the Mn-O-Mn bond angle, which according to
ref\cite{alonso_new} for 2$\%$ decrease in bond length, increases
by less than 1$^o$, the \emph{T$_c$} is found to decrease further
by 7 K \cite{commentpr}. From analysis of
electronic structure, upon application of pressure, the bond
lengths shorten which in turn enhances the Mn-O and Tl-O$^{'}$
interactions resulting into the broadening of the net band width.
The more important effect for the present issue, however, is that
the delocalized Tl-O$^{'}$ effective levels are found to shift \cite{comnew}. 
As is evident from the level diagram in Fig. \ \ref{level}, the
strength of the hybridization between the magnetic and the
nonmagnetic states depends crucially on the positioning of the
various energy levels. The movement of the energy levels causes
changes in hybridization strengths, leading to suppression in the
kinetically driven magnetic interactions and hence in a reduction
of T$_c$. The small increase of Mn-O-Mn bond angle in the second
set of calculations, which is a secondary effect on top of the
former effect, increases the antiferromagnetic super-exchange
contribution resulting into further suppression of ferromagnetic
T$_c$.

To conclude, we have shown by means of NMTO based {\it
downfolding} study that the underlying mechanism of ferromagnetism
in Tl$_{2}$Mn$_{2}$O$_{7}$ is a kinetic energy driven mechanism originally proposed
for Sr$_{2}$FeMoO$_{6}$. The mean field T$_c$, estimated using
computed $J'$s is in good agreement with experiments. For Tl$_{2}$Mn$_{2}$O$_{7}$
moderately doped with Sb, we found an enhancement in T$_c$ and on
application of pressure, T$_c$ is found to decrease, both in
agreement with experimental observations. The microscopic origin
of these changes in T$_c$ are found to be dominantly due to
changes in the magnetic interaction strengths arising from the proposed
kinetically-driven mechanism.

The research was funded by DST project SR/S2/CMP-42/2003. We thank MPG-partnergroup program for the collaboration.




\begin{thebibliography}{99}
\bibitem{note} Empty Mn-$e_g$ does not contribute in the mechanism.
\end{thebibliography}
\end{document}